\documentclass[twocolumn]{aastex631}

\usepackage{amsmath}

\begin{document}

\title{Extracting the Temperature Analytically in Hydrodynamics Simulations with Gas and Radiation Pressure}

\author[0000-0002-6316-602X]{Thomas W. Baumgarte}
\affiliation{Bowdoin College, Brunswick, Maine 04011, USA }

\author[0000-0002-3263-7386]{Stuart L. Shapiro}
\affiliation{Department of Physics, University of Illinois at Urbana-Champaign, Urbana, Illinois 61801, USA}
\altaffiliation{Department of Astronomy and NCSA, University of Illinois at Urbana-Champaign, Urbana, Illinois 61801, USA}

\begin{abstract}
Numerical hydrodynamics simulations of gases dominated by ideal, nondegenerate matter pressure and thermal radiation pressure in equilibrium entail finding the temperature as part of the evolution.  Since the temperature is not typically a variable that is evolved independently, it must be extracted from the the evolved variables (e.g.~the rest-mass density and specific internal energy). This extraction requires solving a quartic equation, which, in many applications, is done numerically using an iterative root-finding method.  Here we show instead how the equation can be solved analytically and provide explicit expressions for the solution.  We also derive Taylor expansions in limiting regimes and discuss the respective advantages and disadvantages of the iterative vs.~analytic approaches to solving the quartic.
\end{abstract}


\section{Introduction}

The numerical modeling of a vast range of astrophysical systems and scenarios entails solving the equations of either Newtonian or relativistic hydrodynamics. This computation requires integrating the Euler equation (or some generalization) together with the continuity equation and the first law of thermodynamics (see, e.g., \cite{LanL87}).  In many codes, the latter two are integrated forward in time to provide updated values for the rest-mass density $\rho_0$ and the specific internal energy density $e$.  Sometimes these quantities are evolved directly, or in typical high-resolution, shock-capturing schemes they may be extracted from related conserved quantities in a reconstruction step (see, e.g., \cite{Tor09}; see also \cite{BauS10,RezZ13} for treatments in a relativistic context).

Given $\rho_0$ and $e$ one typically computes the pressure $P$, which is required in the Euler equation as well as any thermal heating and/or cooling (i.e.~entropy changing) terms in the energy equation.  In gases that are dominated by ideal, nondegenerate fluid pressure and radiation pressure in equilibrium with matter, i.e.~with matter obeying a Maxwell-Boltzmann pressure law together with blackbody radiation pressure, this entails knowing the temperature $T$ (see Eq.~(\ref{pressure}) below).  The latter can be found by solving a quartic equation in which the coefficients depend on $\rho_0$ and $e$ (see Eq.~(\ref{int_energy}) below).  

In many applications, the quartic equation (\ref{int_energy}) is solved using an iterative numerical root-finding method, e.g.~a Newton-Raphson algorithm (see, e.g., \cite{PreTVF02}), but it does not appear to be widely appreciated that the solution can also be found analytically in a few steps.  The main purpose of this very short paper is to provide explicit expressions for the exact analytical solution (see Section \ref{sec:exact} below), together with rapidly-converging Taylor expansions that yield excellent approximations in two opposite regimes (see Section \ref{sec:approx}).  We conclude in Section \ref{sec:discussion} with a discussion of the relative merits of the analytic and iterative methods for finding the temperature $T$.

\section{Exact Analytical Solutions}
\label{sec:exact}

In matter obeying a Maxwell-Boltzmann distribution with electromagnetic radiation in thermal equilibrium the pressure $P$ is given by
\begin{equation} \label{pressure}
P = \frac{\rho_0 k_B T}{\mu m_B} + \frac{1}{3} a T^4
\end{equation}
and the internal energy per unit mass $e$ by
\begin{equation} \label{int_energy}
e = \frac{3}{2} \frac{k_B T}{\mu m_B} + \frac{a T^4}{\rho_0}.
\end{equation}
Here $k_B$ is the Boltzmann constant, $\mu$ the mean molecular weight, $m_B$ the baryon mass, and $a$ the radiation constant.
An equation of state like the above often applies in the interior of stars, in accretion disks around black holes, and in the coupled baryon-radiation fluid in the early universe, to name a few familiar astrophysical environments.

Given values of $\rho_0$ and $e$, Eq.~(\ref{int_energy}) needs to be solved for the temperature $T$ before the pressure (\ref{pressure}) can be evaluated.  In many numerical simulations this is done with an iterative root-finding method, e.g.~a Newton-Raphson scheme.  It is also possible, however, to solve the equation analytically in just a few steps, as we demonstrate in the following.

We start by writing Eq.~(\ref{int_energy}) in the form
\begin{equation} \label{quartic_1}
    T^4 + \beta^3 T - \gamma^4 = 0
\end{equation}
where we have defined the coefficients
\begin{subequations} \label{beta_and_gamma}
\begin{align}
    \beta^3 & \equiv \frac{3}{2}\frac{\rho_0 k_B}{\mu m_B a} > 0 \\
    \gamma^4 & \equiv \frac{e \rho_0}{a} > 0
\end{align}
\end{subequations}
so that both $\beta$ and $\gamma$ have units of temperature.  We next express the quartic Eq.~(\ref{quartic_1}) as a product of two quadratic expressions,
\begin{equation} \label{quartic_2}
T^4 + \beta^3 T - \gamma^4 =
(T^2 + a T + b^2)(T^2 + c T - d^2)
\end{equation}
(see, e.g., \cite{AbrS72}).  We have again chosen the powers of the coefficients $a$, $b$, $c$, and $d$ so that they have units of temperature, but note that they may not end up being real.  Once these coefficients have been identified, the four roots for $T$ are then given by
\begin{subequations} \label{roots}
\begin{align}
T_{1,2} & = - \frac{a}{2} \pm \left( \frac{a^2}{4} - b^2 \right)^{1/2}, \label{roots12} \\
T_{3,4} & = - \frac{c}{2} \pm \left( \frac{c^2}{4} + d^2 \right)^{1/2}, \label{roots34}
\end{align}
\end{subequations}
where the subscripts label the four different roots.

In order to find the four coefficients $a$, $b$, $c$, and $d$ we expand the right-hand side of (\ref{quartic_2}) and match corresponding powers of $T$ on both sides of the equation to find the four expressions
\begin{subequations} \label{coefs}
\begin{align}
a + c & = 0, \label{coef1} \\
b^2 - d^2 + ac & = 0, \label{coef2}\\
c b^2 - a d^2 & = \beta^3, \label{coef3}\\
b^2 d^2 & = \gamma^4.\label{coef4}
\end{align}
\end{subequations}
From (\ref{coef1}) we evidently have $c = -a$, which we can use to write Eqs.~(\ref{coef2}) and (\ref{coef3}) as 
\begin{subequations}
\begin{align}
b^2 - d^2 & = a^2, \\
b^2 + d^2 & = - \beta^3/a. 
\end{align}
\end{subequations}
Adding and subtracting the two equations then yields equations for $b^2$ and $d^2$ in terms of $a$ only, namely 
\begin{subequations} \label{b_and_d}
\begin{align}
b^2 & = -\frac{\beta^3}{2a} + \frac{a^2}{2}, \\
d^2 & = -\frac{\beta^3}{2a} - \frac{a^2}{2}.
\end{align}
\end{subequations}
Inserting these into (\ref{coef4}) we then find the cubic equation
\begin{equation} \label{cubic}
a^6 + 4 \gamma^4 a^2 - \beta^6 = 0.
\end{equation}
for $a^2$.  

We next solve (\ref{cubic}) analytically using the recipe on page~228 of \cite{PreTVF02}.  For convenience and easier comparison with \cite{PreTVF02} we rewrite (\ref{cubic}) as 
\begin{equation} \label{cubic_2}
x^3 + \bar a x^2 + \bar b x + \bar c = 0
\end{equation}
where $x = a^2$, $\bar a = 0$, $\bar b = 4 \gamma^4 > 0$, and $\bar c = - \beta^6 < 0$.  We next compute the two quantities
\begin{subequations} \label{Q_and_R}
\begin{align}
Q & = \frac{\bar a^2 - 3 \bar b}{9} = - \frac{4}{3} \gamma^4 < 0,\\
R & = \frac{2 \bar a^3 - 9 \bar a \bar b + 27 \bar c}{54} = - \frac{\beta^6}{2} < 0.
\end{align}
\end{subequations}
Since both $R$ and $Q$ are negative we have $R^2 > Q^3$. In this case there exists only one real solution, given by
\begin{subequations} \label{x}
\begin{equation} \label{x_1}
x = A + B,
\end{equation}
where
\begin{align} \label{x_A}
A & \equiv \left( |R| + \sqrt{R^2 - Q^3} \right)^{1/3} \nonumber \\
& = \left( \frac{\beta^6}{2} + \left( \frac{\beta^{12}}{4} + \frac{4^3 \gamma^{12}}{3^3} \right)^{1/2} \right)^{1/3} > 0
\end{align} 
and
\begin{equation}  \label{x_B}
B = \frac{Q}{A} < 0.
\end{equation}
\end{subequations}
Even though the two terms in (\ref{x_1}) have opposite signs, $x$ is positive for all positive $\beta$ and $\gamma$.  

We can now compute the roots $T$ of (\ref{int_energy}) for given values of $e$ and $\rho_0$ as follows.  We first identify $\beta$ and $\gamma$ from (\ref{beta_and_gamma}).  Given these, we compute the real root $x$ to the cubic equation (\ref{cubic_2}) from (\ref{x}) and take its positive square root to obtain the coefficient $a = x^{1/2} > 0$.  Knowing $a$ we immediately have $c = -a$ and can find $b^2$ and $d^2$ from (\ref{b_and_d}).  Given the four coefficients $a$, $b$, $c$, and $d$ we can compute the four roots to the quartic equation (\ref{quartic_2}) from (\ref{roots}).  Note, however, that the discriminant in the roots (\ref{roots34}) is always negative, 
\begin{equation}
\frac{c^2}{4} + d^2 = - \frac{\beta^3}{2a} - \frac{a^2}{4} < 0, 
\end{equation}
so that only (\ref{roots12}) can provide real roots.  The desired positive real root $T$ is then given by the ``+" sign,
\begin{equation} \label{T_exact}
T = - \frac{a}{2} + \left( \frac{a^2}{4} - b^2 \right)^{1/2} 
= - \frac{a}{2} + \left( \frac{\beta^3}{2 a} - \frac{a^2}{4} \right)^{1/2},
\end{equation}
providing us with an exact, analytic solution to the quartic equation (\ref{int_energy}).

\section{Approximate Analytical Solutions}
\label{sec:approx}

Despite the appeal of an exact analytic solution, a numerical value computed from (\ref{T_exact}) may actually be less accurate than one obtained from an iterative root-finding method. Specifically, when $\beta \ll \gamma$, the two coefficients $A$ and $B$ in (\ref{x}) end up with very similar magnitude but opposite sign, so that round-off error can lead to a large numerical error.  To see this, we write
\begin{equation}
    \beta = \epsilon \gamma,
\end{equation}
where the dimensionless parameter $\epsilon$ can be computed from (\ref{beta_and_gamma}) as
\begin{equation} \label{epsilon}
\epsilon = \frac{\beta}{\gamma} = \frac{3^{1/3}}{2^{1/3} \mu^{1/3}} \frac{\rho_0^{1/12} k_B^{1/3}}{m_B^{1/3} a^{1/12} e^{1/4}}.
\end{equation}
Assuming that $\epsilon$ is sufficiently small we use a Taylor expansion to find
\begin{align}
A & = \frac{2}{3^{1/2}} \gamma^2 + \frac{1}{8} \epsilon^6 \gamma^2 + \mathcal{O}(\epsilon^{12}) \\
B & = - \frac{2}{3^{1/2}} \gamma^2 + \frac{1}{8} \epsilon^6 \gamma^2 + \mathcal{O}(\epsilon^{12}) 
\end{align}
Even for moderate ratios of $\epsilon \simeq 10^{-2}$, say, 
numerically evaluating the sum $A + B$ in (\ref{x_1}) with double precision will lead to large round-off error, rendering the exact analytical solution unreliable in this regime.  

\begin{figure}[t]
    \centering
    \includegraphics[width=0.45\textwidth]{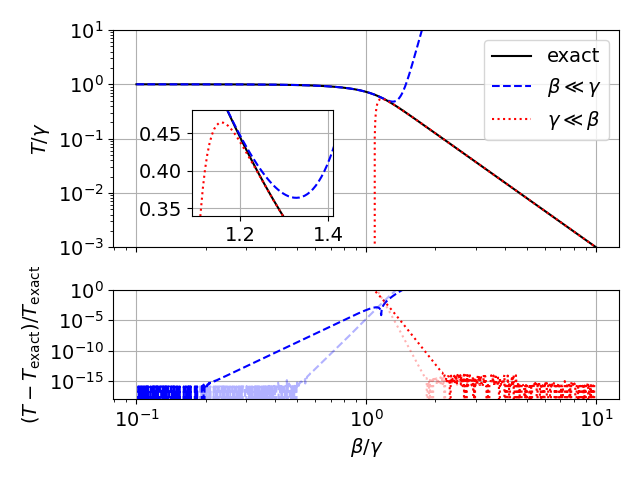}
    \caption{A comparison of the exact temperature (in units of $\gamma$) with two expansions in opposite limits of $\beta / \gamma$.  The (black) solid line shows the exact value (\ref{T_exact}), the (blue) dashed line the expansion (\ref{T_expand_1}) for $\beta \ll \gamma$ and the (red) dotted line the expansion (\ref{T_expand_2}) for $\beta \gg \gamma$.  Evidently, the two expansions show excellent agreement with the exact value except in a small region around $\beta / \gamma \simeq 1.2$.  In the bottom panel we show the relative errors between exact and expanded values. The dark-colored lines show results using the (black) leading-order terms in the top lines in expansions (\ref{T_expand_1}) and (\ref{T_expand_2}) only, while the lighter-colored lines include the additional (gray) terms in the remaining lines. }
    \label{fig:T_comparison}
\end{figure}

However, we may also express the exact temperature (\ref{T_exact}) in terms of a Taylor expansion, which yields 
\begin{align} \label{T_expand_1}
    T = \gamma \Big( & 1 - \frac{1}{4} \epsilon^3 
    - \frac{1}{32} \epsilon^6 
    + \frac{7}{2048} \epsilon^{12} 
    + \frac{1}{512} \epsilon^{15} \nonumber \\
    & {\color{gray}+ \frac{39}{65536} \epsilon^{18}
    - \frac{1045}{8388608} \epsilon^{24} 
    - \frac{11}{131072} \epsilon^{27} }\\ 
    & {\color{gray}- \frac{7735}{268435456} \epsilon^{30}} 
    +  \mathcal{O}(\epsilon^{36}) \Big) 
    \nonumber \\ 
    & \hspace{1.8in} (\epsilon = \beta/\gamma < 1). \nonumber
\end{align}
Here the lighter gray terms in the second and third line are higher-order terms that are included in the lighter-colored lines in the bottom panel of Fig.~\ref{fig:T_comparison}.
Remarkably, even for $\epsilon \simeq 10^{-1}$ the first four terms are sufficient to compute the exact temperature to machine precision in double precision applications.

We may also consider the opposite limit, i.e.~$\beta \gg \gamma$ or $\epsilon \gg 1$.  In this case a Taylor expansion about $\delta \equiv \epsilon^{-1} = 0$ yields 
\begin{align} \label{T_expand_2}
T = \beta \Big( & \delta^4 - \delta^{16} 
+ 4 \delta^{28} - 22 \delta^{40} \nonumber \\ 
& {\color{gray}+ 140 \delta^{52} - 969 \delta^{64}} + \mathcal{O}(\delta^{76}) \Big) \\
& \hspace{1.8in} (\delta = \gamma/\beta< 1), \nonumber 
\end{align}
providing an even faster converging series than (\ref{T_expand_1}).  It is easy to extend both expansions (\ref{T_expand_1}) and (\ref{T_expand_2}) to higher order if desired.

In Fig.~\ref{fig:T_comparison} we compare the exact values (\ref{T_exact}) for the temperature with the expanded values (\ref{T_expand_1}) and (\ref{T_expand_2}). Specifically, we divide Eq.~(\ref{quartic_1}) by $\gamma^4$ to obtain
\begin{equation}
\left(\frac{T}{\gamma}\right)^4 + \left(\frac{\beta}{\gamma}\right)^3 \left(\frac{T}{\gamma}\right) - 1 = 0,
\end{equation}
so that $T/\gamma$, which is shown in Fig.~\ref{fig:T_comparison}, depends on $\beta/\gamma$ alone.  Evidently, the expanded values provide excellent estimates for the temperature, with relative deviations exceeding double-precision accuracy only in a small region around $\beta / \gamma \simeq 1$, and even in this region the relative error never exceeds about 1\%.  The size of the region depends on the number of terms included in the expansion, as demonstrated in the lower panel of Fig.~\ref{fig:T_comparison}.   We note that we used iteratively computed values for the temperature $T_{\rm exact}$ in lower panel of Fig.~\ref{fig:T_comparison}, since otherwise the error would have been dominated by the round-off error in (\ref{T_exact}) for small values of $\beta / \gamma$, as we discussed at the beginning of this Section.

\section{Discussion}
\label{sec:discussion}

The quartic equation (\ref{int_energy}) applies in many different astrophysical environments -- we already mentioned the interior of stars, accretion disks, and the early Universe -- and its solution is therefore required in a variety of different numerical approaches and approximations that treat Newtonian or general relativistic hydrodynamics  or magnetohydrodynamics in which thermal gas and radiation pressure are both present. Whether or not using any of the exact analytical solutions presented above is preferable to using a Newton-Raphson iterative method is therefore likely to depend on the context.

\begin{table}[!t]
    \centering
    \begin{tabular}{c|c|c|c}
         star&  $\rho_{0c}$ [g/cm$^3$] & $T_c$ [K] & $\beta / \gamma$ \\
         \hline
        1 $M_\odot$ main sequence& $7.7 \times 10^1$ & $1.2 \times 10^7$
        & 1.84 \\
        1.3 $M_\odot$ red giant  & $3.5 \times 10^5$ & $4.0 \times 10^8$ & 2.74 \\
        $10^6 M_\odot$ supermassive  & $10^{-3}$ & $2\times 10^7$ & 0.16 
    \end{tabular}
    \caption{Representative values for the density $\rho_0$ and temperature $T$ at the center of a (zero-age) main-sequence and a red giant star (see Tables 28.3 and 28.7 in \cite{Sch58}), and a (hypothetical) supermassive star (see \cite{ShaT83}).  For each example we first compute $e$ from (\ref{int_energy}), and then evaluate the ratio $\beta/\gamma$ from (\ref{beta_and_gamma}).}
    \label{tab:example}
\end{table}

In terms of computational speed, we note that evaluating the exact analytic temperature (\ref{T_exact}) entails taking square and cube roots in Eq.~(\ref{x_A}), which is not necessary when using a root-finding method, e.g.~a Newton-Raphson algorithm. One might therefore expect that using the latter with a sufficiently good initial guess will generally be faster than the former.  We confirmed this expectation in some experiments using {\tt C++}, even though the specific results will depend on the compiler, the numerical platform, and on how exactly these roots are evaluated.   In our experiments using {\tt python}, on the other hand, we found that evaluating the exact analytical temperature was faster than the iterative method when using the {\tt math} library's {\tt sqrt} and {\tt cbrt} commands for the roots (or the {\tt **(1./3.)} or {\tt **0.5} operations), but slower when using the {\tt numpy} library.  We again assume that the specific results will depend on the version of the libraries and the computational platform.  Even though the approximate analytical expressions (\ref{T_expand_1}) and (\ref{T_expand_2}) also require taking roots (namely when finding $\beta$ and $\gamma$ from (\ref{beta_and_gamma})), their evaluation was in general either similar in speed to the iterative methods in our experiments with {\tt C++}, or slightly faster.   Whether or not these considerations are important again depends on the context:  in simulations in which extracting the temperature takes a significant fraction of the entire simulation the speed of the extraction is evidently quite important.  In larger multiphysics codes, on the other hand, the time needed for the extraction may be sub-dominant, making other considerations more important.

One such consideration is the fact that all of the above analytic expressions avoid the need for an initial guess.  In many numerical evolution calculations one may adopt the previous value of the temperature as the initial guess, but this requires extra storage for the previous temperature.  Alternatively, one might use the temperature at a neighboring grid-point as the initial guess -- but that means that, in parallel computations, the results obtained with different number of processors, when grid-points are evaluated in different order, will not agree to machine-precision.  Moreover, iterative methods may not automatically converge to the correct solution, while the analytic approach of Section \ref{sec:exact} identifies the unique desired solution without ambiguity.  In practice, one might also consider using an analytical expression to obtain a reliable and accurate initial guess for the iterative method without the need for additional storage, or adopting different approaches in different regimes of $\beta / \gamma$.

As specific examples of astrophysical environments we list representative values for the central density $\rho_0$ and temperature $T$ for three different types of stars in Table \ref{tab:example}.  From these values we compute the specific internal energy $e$ from (\ref{int_energy}), and $\beta$ and $\gamma$ from (\ref{beta_and_gamma}).  Although the densities $\rho_0$ in particular differ quite dramatically for the different examples, the range in the ratios $\beta / \gamma$ is significantly smaller, as expected from (\ref{epsilon}).  For these examples, the round-off error in the exact solution (\ref{T_exact}) should only be moderate, while the expanded solutions (\ref{T_expand_1}) and (\ref{T_expand_2}) should provide excellent approximations.

\hspace{0.1in}

This work was supported in parts by National Science Foundation (NSF) grant PHY-2308821 to Bowdoin College, as well as NSF grants PHY-2006066 and PHY-2308242 to the University of Illinois at Urbana-Champaign.

\software{{\tt python}, including the {\tt math} library \citep{python}, {\tt numpy} \citep{numpy}, {\tt matplotlib} \citep{matplotlib}.}


\begin{thebibliography}{}
\expandafter\ifx\csname natexlab\endcsname\relax\def\natexlab#1{#1}\fi
\providecommand{\url}[1]{\href{#1}{#1}}
\providecommand{\dodoi}[1]{doi:~\href{http://doi.org/#1}{\nolinkurl{#1}}}
\providecommand{\doeprint}[1]{\href{http://ascl.net/#1}{\nolinkurl{http://ascl.net/#1}}}
\providecommand{\doarXiv}[1]{\href{https://arxiv.org/abs/#1}{\nolinkurl{https://arxiv.org/abs/#1}}}

\bibitem[{{Abramowitz} \& {Stegun}(1972)}]{AbrS72}
{Abramowitz}, M., \& {Stegun}, I.~A. 1972, {Handbook of Mathematical Functions}
  (Dover)

\bibitem[{{Baumgarte} \& {Shapiro}(2010)}]{BauS10}
{Baumgarte}, T.~W., \& {Shapiro}, S.~L. 2010, {Numerical Relativity: Solving
  Einstein's Equations on the Computer} ({Cambridge University Press})

\bibitem[{Harris {et~al.}(2020)Harris, Millman, van~der Walt, Gommers,
  Virtanen, Cournapeau, Wieser, Taylor, Berg, Smith, Kern, Picus, Hoyer, van
  Kerkwijk, Brett, Haldane, del R{\'{i}}o, Wiebe, Peterson,
  G{\'{e}}rard-Marchant, Sheppard, Reddy, Weckesser, Abbasi, Gohlke, \&
  Oliphant}]{numpy}
Harris, C.~R., Millman, K.~J., van~der Walt, S.~J., {et~al.} 2020, Nature, 585,
  357, \dodoi{10.1038/s41586-020-2649-2}

\bibitem[{Hunter(2007)}]{matplotlib}
Hunter, J.~D. 2007, Computing in Science \& Engineering, 9, 90,
  \dodoi{10.1109/MCSE.2007.55}

\bibitem[{{Landau} \& {Lifshitz}(1987)}]{LanL87}
{Landau}, L.~D., \& {Lifshitz}, E.~M. 1987, {Fluid Mechanics} ({Butterworth
  Heinemann})

\bibitem[{{Press} {et~al.}(2002){Press}, {Teukolsky}, {Vetterling}, \&
  {Flannery}}]{PreTVF02}
{Press}, W.~H., {Teukolsky}, S.~A., {Vetterling}, W.~T., \& {Flannery}, B.~P.
  2002, {Numerical Recipes in C++: the Art of Scientific Computing} ({Cambridge
  University Press})

\bibitem[{{Rezzolla} \& {Zanotti}(2013)}]{RezZ13}
{Rezzolla}, L., \& {Zanotti}, O. 2013, {Relativistic Hydrodynamics} ({Oxford
  University Press})

\bibitem[{{Schwarzschild}(1958)}]{Sch58}
{Schwarzschild}, M. 1958, {Structure and Evolution of the Stars} (Dover)

\bibitem[{{Shapiro} \& {Teukolsky}(1983)}]{ShaT83}
{Shapiro}, S.~L., \& {Teukolsky}, S.~A. 1983, {Black holes, white dwarfs and
  neutron stars. The physics of compact objects} (Wiley),
  \dodoi{10.1002/9783527617661}

\bibitem[{{Toro}(2009)}]{Tor09}
{Toro}, E.~F. 2009, {Riemann Solvers and Numerical Methods for Fluid Dynamics}
  (Springer)

\bibitem[{Van~Rossum(2020)}]{python}
Van~Rossum, G. 2020, The Python Library Reference, release 3.8.2 (Python
  Software Foundation)

\end{thebibliography}
\bibliographystyle{aasjournal}

\end{document}